\documentclass[%
 reprint,
 amsmath,amssymb,
 aps,
 prl,
 superscriptaddress,
]{revtex4-2}

\usepackage{graphicx}
\usepackage{dcolumn}
\usepackage{bm}
\usepackage[hidelinks]{hyperref}
\usepackage{xurl}
\begin{document}

\preprint{APS/123-QED}

\title{Precise measurement of the $\Lambda$-binding energy difference between $^3_\Lambda$H and $^4_\Lambda$H \\via decay-pion spectroscopy at MAMI}

\author{Ryoko~Kino}
\email{ryoko.kino@riken.jp.}
\affiliation{RIKEN Nishina Center for Accelerator-Based Science, Saitama, 351-0198, Japan}

\author{Sho~Nagao}
\email{sho.nagao@nex.phys.s.u-tokyo.ac.jp.}
\affiliation{Department of Physics, Graduate School of Science, The University of Tokyo, Tokyo 113-0033, Japan}
\affiliation{Quark Nuclear Science Institute, Graduate School of Science, The University of Tokyo, Tokyo, 113-0033, Japan}

\author{Patrick~Achenbach}
\affiliation{Institute for Nuclear Physics, Johannes Gutenberg University Mainz, 55099 Mainz, Germany}

\author{Satoshi~N.~Nakamura}
\affiliation{Department of Physics, Graduate School of Science, The University of Tokyo, Tokyo 113-0033, Japan}
\affiliation{Quark Nuclear Science Institute, Graduate School of Science, The University of Tokyo, Tokyo, 113-0033, Japan}
\affiliation{Department of Physics, Graduate School of Science, Tohoku University, Miyagi 980-8578, Japan}

\author{Josef~Pochodzalla}
\affiliation{Institute for Nuclear Physics, Johannes Gutenberg University Mainz, 55099 Mainz, Germany}
\affiliation{Helmholtz Institute Mainz, GSI Helmholtzzentrum f\"ur Schwerionenforschung, Darmstadt, and Johannes Gutenberg University Mainz, 55099 Mainz, Germany}

\author{Takeru~Akiyama}
\affiliation{Department of Physics, Graduate School of Science, Tohoku University, Miyagi 980-8578, Japan}
\affiliation{Graduate Program on Physics for the Universe (GP-PU), Tohoku University, Miyagi 980-8578, Japan}

\author{Ralph~B\"ohm}
\affiliation{Facility for Antiproton and Ion Research (FAIR), 64291 Darmstadt, Germany}

\author{Mirco~Christmann}
\author{Michael~O.~Distler}
\author{Luca~Doria}
\author{Anselm~Esser}
\author{Julian~Geratz}
\author{Christian~Helmel}
\author{Matthias~Hoek}
\affiliation{Institute for Nuclear Physics, Johannes Gutenberg University Mainz, 55099 Mainz, Germany}

\author{Tatsuhiro~Ishige}
\affiliation{Department of Physics, Graduate School of Science, Tohoku University, Miyagi 980-8578, Japan}
\affiliation{Graduate Program on Physics for the Universe (GP-PU), Tohoku University, Miyagi 980-8578, Japan}

\author{Masashi~Kaneta}
\affiliation{Department of Physics, Graduate School of Science, Tohoku University, Miyagi 980-8578, Japan}

\author{Pascal~Klag}
\affiliation{Institute for Nuclear Physics, Johannes Gutenberg University Mainz, 55099 Mainz, Germany}

\author{David Markus}
\author{Harald Merkel}
\affiliation{Institute for Nuclear Physics, Johannes Gutenberg University Mainz, 55099 Mainz, Germany}

\author{Masaya Mizuno}
\affiliation{Department of Physics, Graduate School of Science, Tohoku University, Miyagi 980-8578, Japan}

\author{Ulrich M\"uller}
\affiliation{Institute for Nuclear Physics, Johannes Gutenberg University Mainz, 55099 Mainz, Germany}

\author{Kotaro~Nishi}
\author{Ken~Nishida}
\affiliation{Department of Physics, Graduate School of Science, The University of Tokyo, Tokyo 113-0033, Japan}

\author{Kazuki~Okuyama}
\affiliation{Department of Physics, Graduate School of Science, Tohoku University, Miyagi 980-8578, Japan}
\affiliation{Graduate Program on Physics for the Universe (GP-PU), Tohoku University, Miyagi 980-8578, Japan}

\author{Jonas~P\"atschke}
\author{Bj\"orn~S\"oren~Schlimme}
\author{Concettina~Sfienti}
\affiliation{Institute for Nuclear Physics, Johannes Gutenberg University Mainz, 55099 Mainz, Germany}

\author{Tianhao~Shao}
\affiliation{Key Laboratory of Nuclear Physics and Ion-beam Application (MOE), Institute of Modern Physics, Fudan University, 200433 Shanghai, China}
\affiliation{Institute for Nuclear Physics, Johannes Gutenberg University Mainz, 55099 Mainz, Germany}

\author{Daniel~Steger}
\author{Marcell~Steinen}
\affiliation{Institute for Nuclear Physics, Johannes Gutenberg University Mainz, 55099 Mainz, Germany}

\author{Liguang~Tang}
\affiliation{Department of Physics, Hampton University, Hampton, Virginia 23668, USA}
\affiliation{Thomas Jefferson National Accelerator Facility, Newport News, Virginia 23606, USA}

\author{Michaela~Thiel}
\author{Philipp~Vonwirth}
\affiliation{Institute for Nuclear Physics, Johannes Gutenberg University Mainz, 55099 Mainz, Germany}

\author{Luca~Wilhelm}
\affiliation{Institute for Nuclear Physics, Johannes Gutenberg University Mainz, 55099 Mainz, Germany}

\collaboration{A1 Collaboration}\noaffiliation

\date{\today}

\begin{abstract}
We performed high-precision decay-pion spectroscopy of light $\Lambda$ hypernuclei at the Mainz Microtron (MAMI) using the A1 spectrometer facility.  
By measuring the monochromatic $\pi^-$ momentum from the two-body weak decay $^3_\Lambda\mathrm{H} \to {}^3\mathrm{He} + \pi^-$ and referencing it to the $^4_\Lambda\mathrm{H} \to {}^4\mathrm{He} + \pi^-$ decay, we determined the $\Lambda$ binding energy of $^3_\Lambda\mathrm{H}$ with unprecedented accuracy.  
The obtained value, $B_\Lambda(^3_\Lambda\mathrm{H}) = 0.523 \pm 0.013_\mathrm{stat.} \pm 0.075_\mathrm{syst.}$~MeV, is consistent with the STAR result, but indicates a significantly deeper binding than inferred from earlier measurements.
This result implies a stronger $\Lambda$-deuteron interaction and provides stringent constraints on hyperon-nucleon interactions.  
\end{abstract}

\maketitle



\section{Introduction}
Understanding the baryon--baryon interaction in terms of its underlying quark degrees of freedom remains a central challenge in modern nuclear physics, owing to the nonperturbative nature of quantum chromodynamics at low energies.
Hypernuclei---nuclei that include one or more hyperons---serve as unique probes of the baryon--baryon interaction beyond the conventional $NN$ sector, providing access to hyperon--nucleon ($YN$) and hyperon--hyperon ($YY$) forces that are otherwise difficult to study experimentally due to the short lifetimes of free hyperons.
High-precision spectroscopy of light hypernuclei has thus become an essential tool for constraining realistic $YN$ potentials, investigating three-body forces, and exploring the flavor dependence of the short-range repulsive core of the nuclear force.

Among light hypernuclei, the hypertriton ($^3_{\Lambda}\mathrm{H}$) and the hypothetical $nn\Lambda$ system occupy central positions.
The hypertriton, a bound three-body $(pn\Lambda)$ system with a small $\Lambda$ binding energy, provides a stringent benchmark for testing $YN$ interactions and constraining the $\Lambda$--$\Sigma$ coupling in few-body calculations~\cite{Fujiwara2007, Hiyama2001}.
Its weak binding and dominant $S$-wave structure make it exquisitely sensitive to subtle variations in the $\Lambda N$ scattering length as well as to the inclusion of three-body $\Lambda NN$ forces~\cite{Kamada1998, Lonardoni2017}.
In contrast, the $nn\Lambda$ system, despite extensive theoretical investigations~\cite{Garcilazo2015}, has not been observed experimentally, and current few-body and lattice-QCD calculations suggest it may be unbound or only marginally bound~\cite{Haidenbauer2019}.

Recent heavy-ion collision experiments by the STAR and ALICE Collaborations have extended the measurement of the $\Lambda$-separation energy and lifetime of the hypertriton to unprecedented precision.
The result published in 2020 by the STAR Collaboration, $B_{\Lambda}=0.406\pm0.120_{\mathrm{stat}}\pm0.110_{\mathrm{syst}}$~MeV~\cite{STAR2020}, was followed by a result published in 2023 by the ALICE Collaboration, $B_{\Lambda}=0.102\pm0.063_{\mathrm{stat}}\pm0.067_{\mathrm{syst}}$~MeV~\cite{Acharya2023} with a lifetime consistent with the free $\Lambda$ value.
Complementary to collider data, nuclear-emulsion analyses using modern image-processing and machine-learning techniques have re-evaluated historical and new J-PARC~E07 emulsion sheets, yielding
$B_{\Lambda}=0.23\pm0.11_{\mathrm{stat}}\pm0.05_{\mathrm{syst}}$~MeV for $^3_{\Lambda}\mathrm{H}$ and
$B_{\Lambda}=2.25\pm0.10_{\mathrm{stat}}\pm0.06_{\mathrm{syst}}$~MeV for $^4_{\Lambda}\mathrm{H}$~\cite{Kasagi2025}.
Indirectly, the in-flight $(K^-,\pi^0)$ measurement program of J-PARC~E73 also aims to refine $B_{\Lambda}(^3_{\Lambda}\mathrm{H})$ through cross-section ratios~\cite{Akaishi2026}.
These new results, taken together, frame the current status of the long-standing ``hypertriton puzzle''---the discrepancy between the weakly bound structure and the unexpectedly short lifetimes reported in some experiments~\cite{chen2023}.
A recent few-body analysis using $p\Lambda$ and $pp\Lambda$ correlation functions suggests that the inclusion of $\Lambda NN$ three-body forces may lead to a comparatively deeper hypertriton binding energy than traditionally expected~\cite{Garrido2024}.

On the theoretical side, the possibility of excited states of the hypertriton has been explored.
Shell-model and few-body calculations predict that any bound excited level would be extremely shallow because of the weak binding and extended spatial distribution of the $\Lambda$ relative to the deuteron core.
While $\gamma$-ray spectroscopy has established excited-state spectra for heavier $\Lambda$ hypernuclei, a confirmed excited state of $^3_{\Lambda}\mathrm{H}$ remains elusive and thus represents an important open question in hypernuclear physics.

\section{Experiment}

The method employed in this paper determines the $\Lambda$ binding energy from the momentum of the $\pi^-$ emitted in the two-body decay of a stopped $\Lambda$ hypernucleus.  
When a 1.5~GeV electron beam from MAMI-C is incident on the target, a proton inside the nucleus is converted into a $\Lambda$ through the $p(e,e'K^+)\Lambda$ reaction.  
A fraction of the produced $\Lambda$ hyperons becomes bound to the residual nucleus, forming a $\Lambda$ hypernucleus that slows down in the target and eventually comes to rest.  
Only those hypernuclei that decay via the two-body weak decay after stopping in the target emit a monochromatic $\pi^-$, for which the invariant-mass relation
\begin{eqnarray}
 m(^A_\Lambda Z) = \sqrt{m(^A(Z+1))^2 + p_{\pi^-}^2}
 + \sqrt{m_{\pi^-}^2 + p_{\pi^-}^2},
 \label{eq:mass}
\end{eqnarray}
holds.  
Here, $m(^A(Z+1))$, $m_{\pi^-}$, and $p_{\pi^-}$ denote the daughter-nucleus mass, $\pi^-$ mass, and pion momentum, respectively.
In the study of the decay $^3_\Lambda\mathrm{H} \to {}^3\mathrm{He} + \pi^-$, the masses of $^3$He and $\pi^-$ are known with precisions of 2.42~eV~\cite{Audi2003} and 0.18~keV~\cite{PDG2024}, respectively.
When the decay occurs at rest, $p_{\pi^-}$ is monochromatic and its precision directly determines that of the hypernuclear mass.
The $\Lambda$ binding energy is defined as
\begin{eqnarray}
 B_\Lambda = m_{\mathrm{core}} + m_\Lambda - m(^A_\Lambda Z),
 \label{eq:def_blambda}
\end{eqnarray}
where $m_\Lambda$ is the $\Lambda$ mass.  
Thus, measuring $p_{\pi^-}$ alone yields $B_\Lambda$ with high precision.

\begin{figure}[ht]
  \centering
  \includegraphics[width=0.48\textwidth]{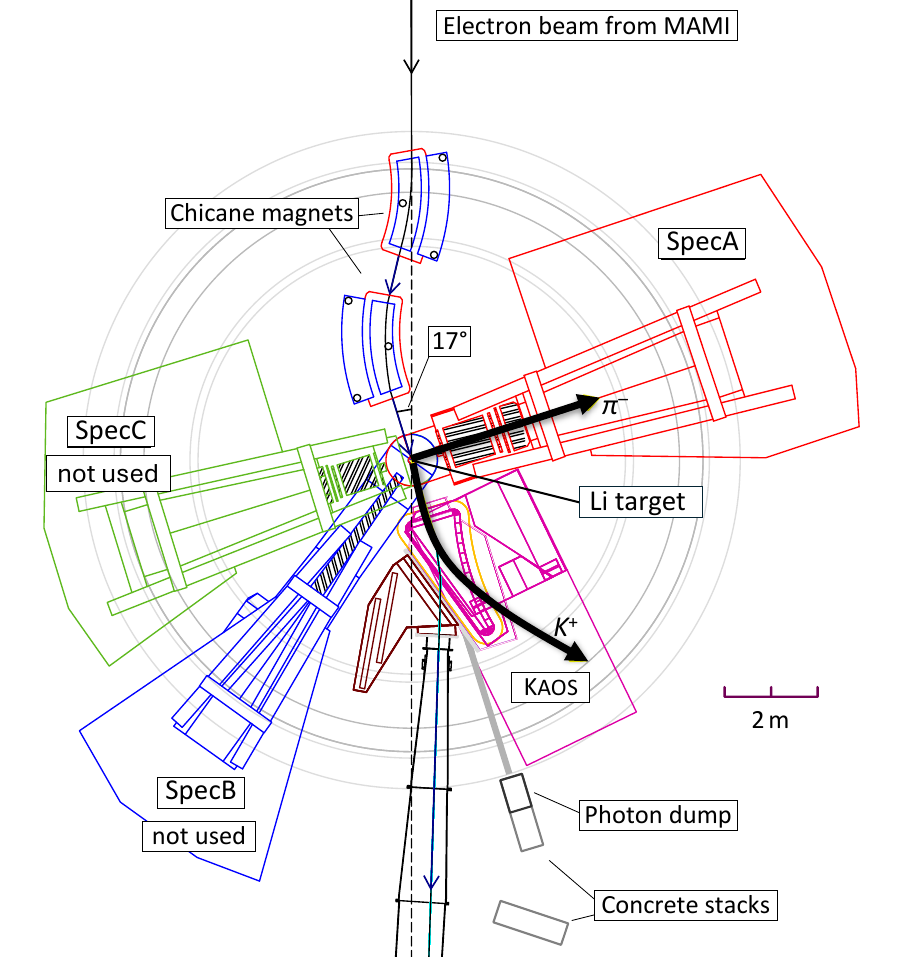}
  \caption{Schematic top view of the experimental setup in the A1 spectrometer facility.}
  \label{fig:A1configulation_for_hyper2022}
\end{figure}

A schematic of the A1 hall is shown in Fig.~\ref{fig:A1configulation_for_hyper2022}.  
Among the high-resolution magnetic spectrometers at A1~\cite{Blomqvist1998},  
Spectrometer~A (SpecA, shown in red) measured the decay $\pi^-$.  
It provides a momentum resolution of $\Delta p/p \sim 2\times10^{-4}$~\cite{Schulz2016NPA}.  
The target chamber and SpecA were connected by a vacuum extension, limiting the total material between target and detectors to the 127~$\mu$m vacuum window.

Strangeness production was tagged by detecting the coincident $K^+$ in the forward spectrometer KAOS, positioned at $0^\circ$~\cite{Senger1993}.  
KAOS consists of a single dipole magnet and a three-layer time-of-flight system for track reconstruction and particle identification.

A natural lithium target was used to reduce possible hyperfragment candidates and electromagnetic background compared to the $^9$Be target used in the previous study~\cite{Schulz2016NPA}.  
To compensate for its low density ($\rho = 0.534$~g/cm$^3$), the target was extended along the beam direction, corresponding to an areal density of 2403~mg/cm$^2$ and enabling high luminosity even at the beam current at 1~$\mu$A.  
The lower current further reduced electromagnetic background.

To minimize energy straggling of decay $\pi^-$, the target width was set to 0.75~mm (40.05~mg/cm$^2$), about 1.25$\sigma$ of the beam width.  
GEANT4 simulations showed that for $\pi^-$ momenta of 110-135~MeV/$c$, that cover the range of relevant decay momenta, the most probable energy loss in the target was only 19.8~keV.

The spectrometer momentum was calibrated using elastic scattering from $^{181}$Ta($e,e')$ and $^{12}$C($e,e')$. 
The momentum of the elastically scattered electron was measured with the same spectrometer SpecA and compared with the precisely calculable kinematic value determined from the beam energy and scattering angle.
The $^{12}$C data were used to calibrate momentum non-linearities along the focal plane via excitation-energy differences, and the $z$-dependence (defined as the position along the beam axis) was corrected using a five-foil configuration spaced 15~mm apart, matching the 45~mm length of the lithium target.  
The dipole field, monitored by NMR, was stable within $\Delta p/p < 10^{-5}$.  
In parallel, beam-energy measurements using undulator radiation interferometry~\cite{Klag2018} were performed at 180-210~MeV to provide an independent calibration reference.
In this work, the absolute momentum scale is referenced to the previously measured $^4_\Lambda\mathrm{H}$ decay $\pi^-$ momentum from earlier MAMI experiments.

The production run lasted 24~days between September and October 2022 with beam currents of $0.5-1.1$~$\mu$A, yielding 287~hours of data and an integrated luminosity of 826.9~fb$^{-1}$, comparable to the previous campaign.
Elastic-scattering runs for calibration were performed in April-May 2024 with beam energies at 180, 195, 210, and 420~MeV, and multiple magnetic field settings per energy.  
The 180-210~MeV beam energies were determined using the undulator method~\cite{Klag2018}, while for 420~MeV the facility’s measured value with $\delta E_b = \pm160$~keV uncertainty~\cite{Schulz2016NPA} was adopted.

\section{Data analysis}

The decay $\pi^-$ momentum in SpecA was reconstructed from the focal-plane positions and angles using the established transfer matrix.  
Event-by-event corrections for energy loss in the target were applied by calculating the path length from the reconstructed emission angle and using the Bethe-Bloch formula.  
Since SpecA is insensitive to the horizontal vertex position ($x$), the path length was calculated by fixing $x$ to the target center.
The uncertainty in the energy loss correction arising from the path-length uncertainty - introduced by the measured beam width of $\sigma = 0.3$~mm and by the spectrometer angle and acceptance - was evaluated using an independent GEANT4 Monte Carlo simulation.
The resulting uncertainty in the most probable energy loss amounts to 16~keV for $p_{\pi^-}\sim114$~MeV/$c$ and it was treated as a systematic error.

\begin{figure*}[ht]
  \centering
  \includegraphics[width=0.7\textwidth]{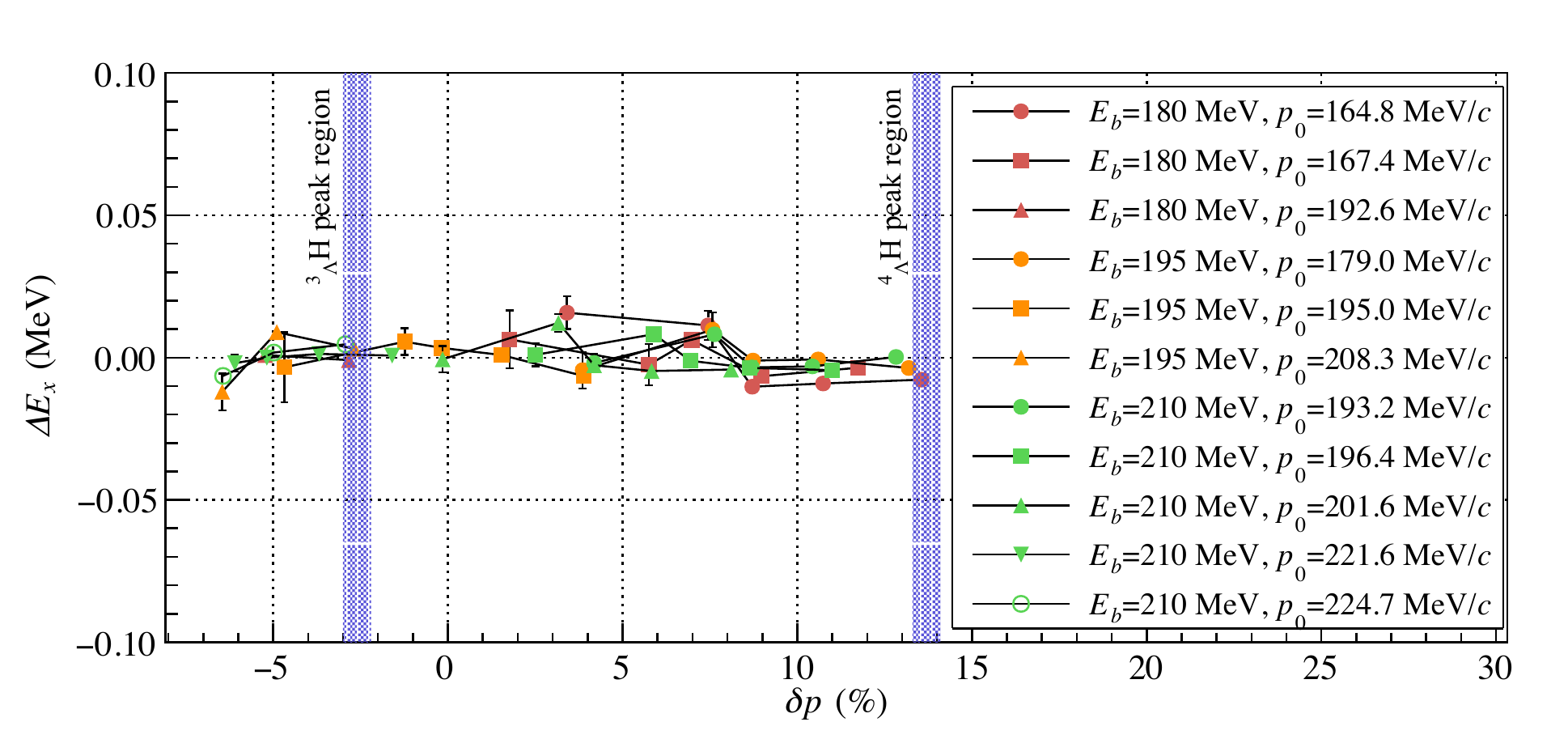}
  \caption{Deviation $\Delta E_x$ from reference excitation energies versus relative momentum $\delta p= (p_{\mathrm{meas.}}/p_0 - 1)\times100$ for the calibration measurements with the $^{12}$C target.}
  \label{fig:linearity}
\end{figure*}

For relative momentum scale calibration, the elastic scattering data were used.  
Since the targets were sufficiently thin, the elastic peaks were fitted with a Landau-Gaussian convolution representing energy loss effects and detector resolution.  
The $^{12}$C data were used to test the linearity of the momentum response by comparing the measured excitation energy spacing with literature values~\cite{ajzenberg1988}. 
The measured excitation energy $E_x^{\mathrm{meas.}}$ for each state was evaluated by calculating the missing mass according to the following equation:
\begin{equation}
\begin{split}
&E_x^{\mathrm{meas.}} \\
=&~ M_{\mathrm{miss}} - M_{^{12}\mathrm{C}} \\
=&~ \sqrt{(E_e + M_{^{12}\mathrm{C}} - E_{e'}^{\mathrm{meas.}})^2 - (\overrightarrow{p}_e - \overrightarrow{p}_{e'}^{\mathrm{meas.}})^2} - M_{^{12}\mathrm{C}}.
\end{split}
\end{equation} 
A small nonlinearity was observed and corrected with a sixth-order polynomial, reducing the residual to $\sigma_{\Delta E_x} \sim 3$~keV.
Figure~\ref{fig:linearity} shows the corrected correlation between the deviation $\Delta E_x = E_x^{\mathrm{ref.}} - E_x^{\mathrm{meas.}}$ and the relative momentum $\delta p = (p_{\mathrm{meas.}}/p_0 - 1)\times100$ across the full acceptance.
Here, $p_{\mathrm{meas.}}$ is the reconstructed momentum of the scattered electron
and $p_0$ is the reference spectrometer momentum setting.
For better comparison, $\Delta E_x$ values are scaled by $195/E_b$, and beam energy offsets are adjusted accordingly in the plot.  
Because the missing mass depends on the scattered-electron momentum as $E_x \simeq \mathrm{const.} - E_{e'}(p)$, the excitation-energy deviation $\Delta E_x$ is linearly proportional to the momentum deviation $\Delta p$ with $dE_x/dp \approx -1$.
Thus, the observed $\sigma_{\Delta E_x} \approx 3$~keV at 195~MeV corresponds to a residual nonlinearity of approximately 2~keV/$c$ at the pion momentum measurement setting ($p_0 = 122$~MeV/$c$).

The $z$-dependence of the reconstructed momentum was evaluated using five-foil $^{181}$Ta targets, for which the scattering-angle uncertainty is smallest.  
A linear correction reduced the residual variation to $\sigma \sim 4$~keV/$c$ over the full $z$ range.

Identification of $K^+$ in KAOS used the measured momentum, $\beta$, and energy loss $dE/dx$.  
True $K^+-\pi^-$ coincidences were selected within a 2~ns gate. 
Accidental events, obtained from 10~ns sidebands of the true coincidence gate, were used to normalize the accidental background in the $\pi^-$ spectra.  
After accounting for these, 39\% of events in the 2~ns gate were estimated to be genuine $K^+-\pi^-$ coincidences; there is also 19\% contamination with $\pi^+-\pi^-$ events due to the finite momentum resolution of KAOS. 

\begin{figure}[ht]
  \centering
  \includegraphics[width=0.48\textwidth]{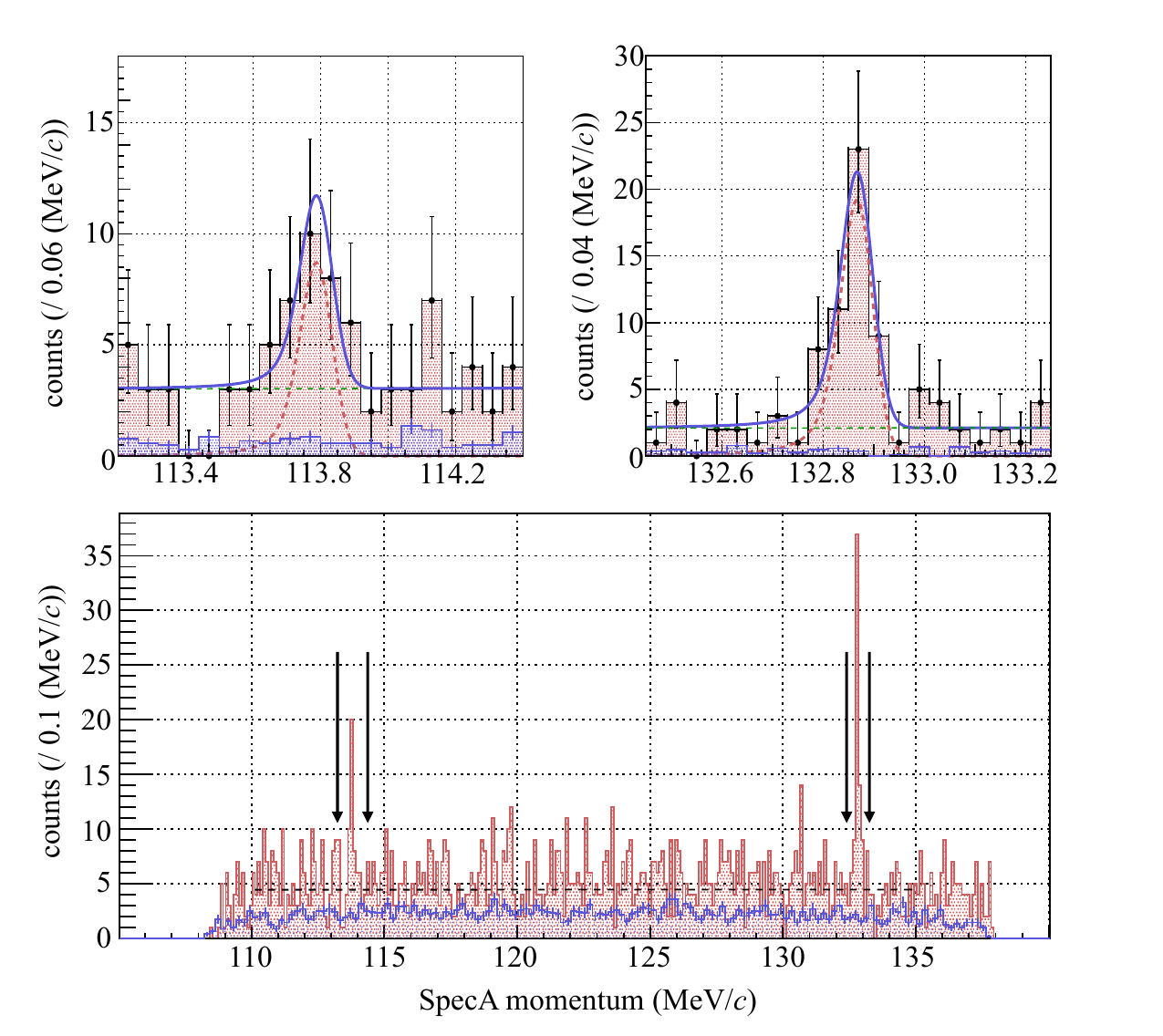}
  \caption{Decay $\pi^-$ momentum distribution measured by SpecA. 
The lower panel shows the spectra for true (red) and accidental (blue) coincidences. 
The upper panels display magnified views of the regions 113.2-114.4~MeV/$c$ and 132.4-133.2~MeV/$c$, together with the results of unbinned fits using a Landau-Gaussian convolution function.}
  \label{fig:pionmomentum}
\end{figure}

Figure~\ref{fig:pionmomentum} shows the $\pi^-$ momentum spectrum within the gate.  
The lower panel presents the true (red) and accidental (blue) spectra, while the upper panel shows unbinned maximum-likelihood fits to the corresponding enlarged spectra.
Two peaks are clearly observed at 114 and 133~MeV/$c$, corresponding to the decays of $^3_\Lambda$H and $^4_\Lambda$H, respectively.

The $^4_\Lambda$H peak was used as the absolute momentum calibration reference; its momentum was determined as $p_{\pi^-}(^4_\Lambda\mathrm{H})=132.631 \pm 0.008$~MeV/$c$.
Compared with the previous result $p_{\pi^-}(^4_\Lambda\mathrm{H})= 132.867 \pm 0.013_\mathrm{stat.} \pm 0.107_\mathrm{syst.}$~MeV/$c$~\cite{Schulz2016NPA}, the calibration factor was $F_{\mathrm{calib}} = 132.867/132.631 = 1.001779$.
After applying the $F_{\mathrm{calib}}$ derived from this peak, the unbinned fit to the $^3_\Lambda$H decay $\pi^-$ yielded 
$p_{\pi^-}(^3_\Lambda\mathrm{H})=113.789 \pm 0.020$~MeV/$c$, as shown in Fig.~\ref{fig:pionmomentum}.

The Landau width $\sigma_L$ and Gaussian width $\sigma_G$ correspond to energy loss and instrumental effects, respectively, and were constrained by simulation and elastic-scattering data.  
Fits with and without width constraints gave consistent peak positions.  
The signal significance, evaluated using the likelihood ratio method~\cite{cousins2008}, was $S_L = \sqrt{-2 \ln (L_{\mathrm{BG}}/{L_{\mathrm{S+BG}})}}=4.33$ for $^3_\Lambda$H, and $S_L=8.36$ for $^4_\Lambda$H, where $L_{\mathrm{BG}}$ and $L_{\mathrm{S+BG}}$ represent the likelihoods under the background-only and signal-plus-background hypotheses.

The systematic uncertainties are dominated by the target energy loss correction and by the calibration reference of the previous $^4_\Lambda$H binding-energy result from MAMI, as summarized in Table~\ref{tab:syst_err}.  
The uncertainty of the target energy loss correction amounts to 17~keV, which consists of a 16~keV contribution from the $x$-position ambiguity in the path-length calculation, evaluated with GEANT4, and an additional $\sim$1~keV uncertainty from the thin-target energy loss treatment in elastic scattering.  
The uncertainty in the referenced $^4_\Lambda$H momentum contributes $108$~keV/$c$~\cite{Schulz2016NPA}. 
Additional effects from non-linearity, $z$ dependence, magnetic-field
stability, and beam position yield a combined uncertainty of $\delta p_{\mathrm{syst.}} = 112$~keV/$c$, corresponding to $\delta B_\Lambda = 75$~keV.

\begin{table*}[t]
\centering
\caption{Summary of the systematic errors.}
\label{tab:syst_err}
\renewcommand{\arraystretch}{1.2}
\begin{tabular}{l|cc}
\hline\hline
Component & $^4_\Lambda$H & $^3_\Lambda$H \\
\hline
Effect of the energy loss within the target (keV)                 & 15  & 17  \\
Stability of the magnetic field (keV/$c$)                 & 1   & 1   \\
Effect of the beam position on momentum reconstruction (keV/$c$) & 1   & 1   \\
Effect of the $z$ position on momentum reconstruction (keV/$c$)  & 4   & 3   \\
Uncertainty in the spectrometer installation angle (keV/$c$)      & $<1$ & $<1$ \\
Momentum non-linearity (keV/$c$)                                  & 2   & 2   \\
Peak position shift due to the peak width (keV/$c$)               & 6   & 6   \\
Uncertainty of referenced $^4_\Lambda$H decay $\pi^-$ momentum (keV/$c$) & 108 & 108 \\
Uncertainty of $\Lambda$ mass (keV/$c^2$)                         & 6   & 6   \\
\hline
Total systematic error in $p_{\pi^-}$ (keV/$c$)                   & 110 & 112 \\
Total systematic error in $B_\Lambda$ (keV)                       & 80  & 75  \\
\hline\hline
\end{tabular}
\end{table*}

\section{Result and Discussion}

The $\Lambda$ binding energy was calculated with Eq.~\ref{eq:def_blambda} using the $^3_\Lambda$H mass obtained from the measured $\pi^-$ momentum via Eq.~\ref{eq:mass}.  
For $^3_\Lambda$H, the core mass $m(^2\mathrm{H})=1875.613~\mathrm{MeV}/c^2$~\cite{Wang2021AME2020} and $m_\Lambda=1115.683~\mathrm{MeV}/c^2$~\cite{PDG2024} were used.  
Using Eq.~\ref{eq:mass} with $m(^3\mathrm{He})=2808.392~\mathrm{MeV}/c^2$~\cite{Wang2021AME2020} and $m_{\pi^-}=139.570~\mathrm{MeV}/c^2$~\cite{PDG2024}, we obtain $m(^3_\Lambda\mathrm{H})=2990.773~\mathrm{MeV}/c^2$.  
The resulting binding energy is
\begin{eqnarray}
 B_\Lambda = 0.523 \pm 0.013_\mathrm{stat.} \pm 0.075_\mathrm{syst.}~\mathrm{MeV}.
 \label{eq:b_lambda}
\end{eqnarray}

A key observable in this work is the difference in decay $\pi^-$ momenta between $^4_\Lambda$H and $^3_\Lambda$H:
\begin{equation}
\begin{split}
&p_{\pi^-}(^4_\Lambda\mathrm{H}) - p_{\pi^-}(^3_\Lambda\mathrm{H}) \\
=&~19.078 \pm 0.022_\mathrm{stat.} \pm 0.036_\mathrm{syst.}~\mathrm{MeV}/c,
\end{split}
\label{eq:momdiff}
\end{equation}
which is uniquely accessible in the present experiment.
Because both decay $\pi^-$ momenta are measured simultaneously within the large momentum acceptance of the same spectrometer and on the same momentum scale, this momentum difference is largely insensitive to the absolute calibration and benefits from a significant cancellation of common systematic uncertainties across the full acceptance.

\begin{figure}[htb]
  \centering
  \includegraphics[width=0.48\textwidth]{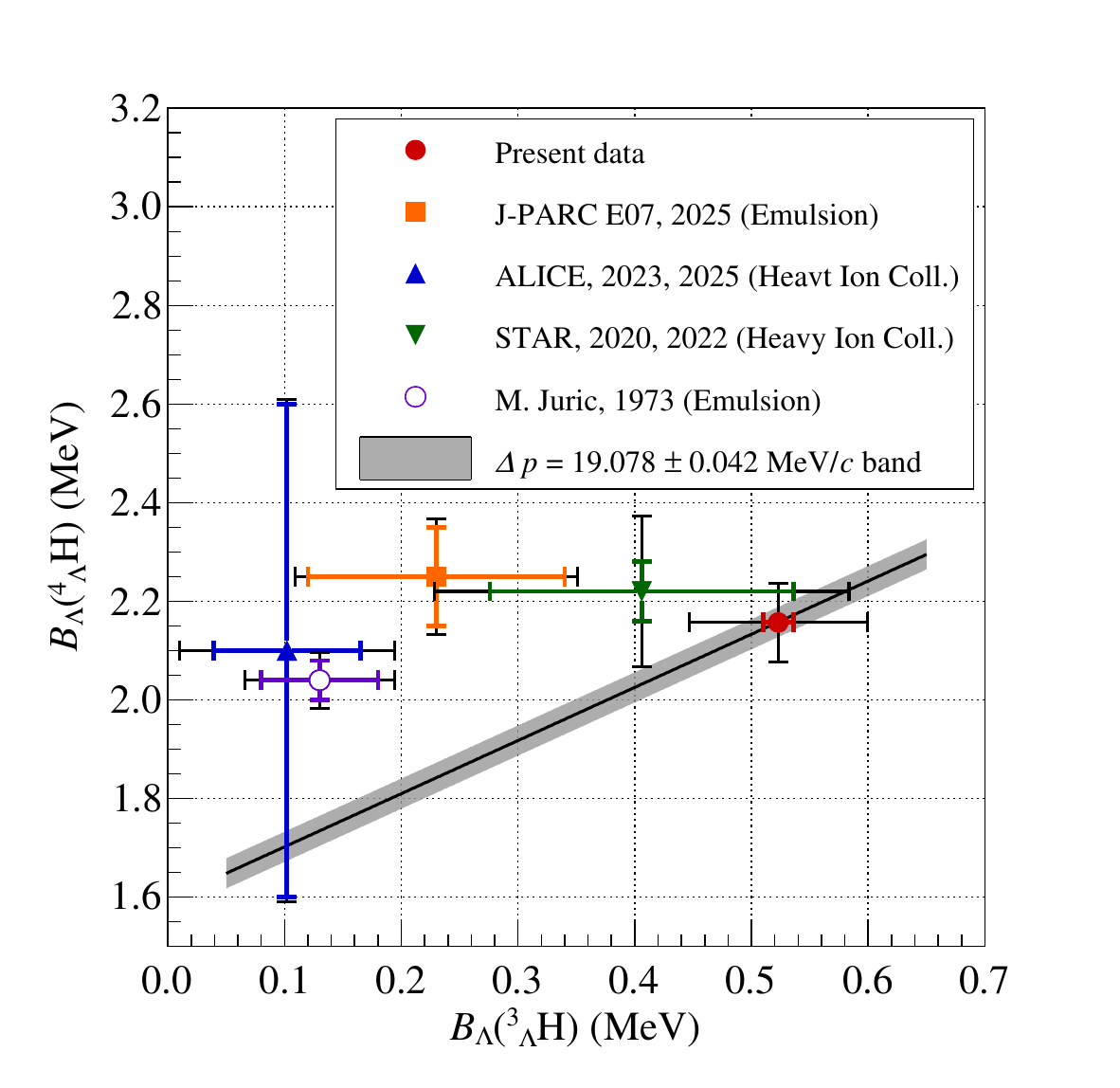}
  \caption{
  Measured difference between $B_\Lambda(^4_\Lambda\mathrm{H})$ and $B_\Lambda(^3_\Lambda\mathrm{H})$ in comparison with independent measurements by J-PARC E07~\cite{Kasagi2025}, ALICE~\cite{Acharya2023,ALICE:2024djx}, STAR~\cite{Shao2022,STAR2020}, and emulsion experiments~\cite{Juric1973}. 
  For the data points, the inner (colored) error bars represent statistical uncertainties only, while the outer error bars show the quadratic sum of statistical and systematic uncertainties.
  For the present work, the black line denotes the central value of the momentum difference, and the shaded band indicates the total uncertainty, calculated as the quadratic sum of the statistical and systematic errors.
  Note that the existing world data for $^3_\Lambda\mathrm{H}$ and $^4_\Lambda\mathrm{H}$ from other experiments are uncorrelated measurements.}
  \label{fig:mom-difference}
\end{figure}

Figure~\ref{fig:mom-difference} shows the correlation between $B_\Lambda(^4_\Lambda\mathrm{H})$ and $B_\Lambda(^3_\Lambda\mathrm{H})$ obtained according to Eq.~\ref{eq:momdiff}.
For comparison, we also plot results from J-PARC E07~\cite{Kasagi2025}, ALICE~\cite{Acharya2023, ALICE:2024djx}, STAR~\cite{Shao2022,STAR2020}, and emulsion measurements~\cite{Juric1973}.  
The curve indicates the present constraint from $p_{\pi^-}(^4_\Lambda\mathrm{H})-p_{\pi^-}(^3_\Lambda\mathrm{H})$, with the shaded band showing its uncertainty.  
Thus, any pair from this measurement $\big(B_\Lambda(^4_\Lambda\mathrm{H}),\,B_\Lambda(^3_\Lambda\mathrm{H})\big)$ must lie on this curve.  
Assuming $B_\Lambda(^4_\Lambda\mathrm{H})=2.17$~MeV from global trends~\cite{eckert2021chart}, our result suggests that $B_\Lambda(^3_\Lambda\mathrm{H})$ is larger than $0.5$~MeV.

Within uncertainties, our value agrees with the STAR measurement~\cite{STAR2020}, but deviates from the earlier emulsion measurements~\cite{Juric1973}. 
However, the emulsion value is a combination of $B_\Lambda(^3_\Lambda\mathrm{H}) = 0.06\pm0.11_\mathrm{stat.}$~MeV from 176 two-body decays and $B_\Lambda(^3_\Lambda\mathrm{H}) = 0.23 \pm 0.11~_\mathrm{stat.}$~MeV  from 46 three-body decays. The reported $B_\Lambda(^4_\Lambda\mathrm{H})$ value from 155 three-body decays is $2.04 \pm 0.04_\mathrm{stat.}$~MeV, while the two-body decay mode is not reported because of larger uncertainties in the $\pi^-$ range-energy relation for $\pi^-$ ranges greater than 3~cm. 
$B_\Lambda(^4_\Lambda\mathrm{H})$ values from 760 two-body decays were reported in~\cite{Bohm1968, Gajewski1967} to be $2.28 \pm 0.11_\mathrm{stat.}$~MeV. 
The apparent systematic differences of about 0.2~MeV in binding energies evaluated from two- and three-body decays may be attributed to a systematic error in the range--energy relation used in emulsion analysis~\cite{bohm1970investigation}. 
According to D.H. Davis~\cite{DAVIS1992369}, a systematic error of 40~keV should be assumed for any emulsion measurement done before 1990. Whether a correlation exists between the systematic errors in the emulsion data for the two hyperhydrogen isotopes remains an open question.

On the theory side, chiral effective field theory calculations~\cite{le2020} show that a deeper $B_\Lambda(^3_\Lambda\mathrm{H})$ can be reproduced by increasing the singlet $^1S_0$ $\Lambda p$ scattering length while weakening the triplet $^3S_1$ term, without contradicting existing $\Lambda p$ scattering data.  
Few-body calculations based on realistic YN interactions and taking channel couplings into account~\cite{Hiyama2014} further indicate that such deeper binding could even entail a bound $nn\Lambda$ system, emphasizing the need for a precise $B_\Lambda(^3_\Lambda\mathrm{H})$ value.

In this work, we successfully measured the decay $\pi^-$ momentum of $^3_\Lambda$H by employing a new $^{\mathrm{nat}}$Li target. 
The momentum scale and reaction‐vertex dependence of the spectrometers were calibrated to a level better than a few keV$/c$ using elastic scattering data obtained with $^{12}$C and $^{181}$Ta targets. 
After that, the decay $\pi^-$ momentum of $^4_\Lambda\mathrm{H}$ previously measured at MAMI~\cite{Schulz2016NPA} was used as an absolute calibration reference, allowing us to determine the relative decay $\pi^-$ momenta of $^4_\Lambda$H and $^3_\Lambda$H.
Our result indicates a deeply bound $\Lambda$ state in $^3_\Lambda$H.

\section*{Acknowledgements}
The authors wish to acknowledge the outstanding support of the accelerator group and operators of MAMI.

This work was funded by the Deutsche Forschungsgemeinschaft (DFG, German Research Foundation) under Grant No.~PO256/7-1 and by the European Union’s Horizon 2020 Research and Innovation Programme under Grant Agreement No.~824093, and the Federal State of Rhineland-Palatinate.  
It was further supported in part by JSPS KAKENHI Grants JP18H05459, JP20H01926, JP24H00219, JP24K00657 and JP25K24464, as well as by the Grant-in-Aid for JSPS Fellows (Grant No.~23KJ0180).
Support from the National Key Research and Development Program of China (Contract No. 2022YFA1604900) and the National Natural Science Foundation of China (Contract Nos. 12025501 and 12147101) is also gratefully acknowledged.

R.K. and K.N. gratefully acknowledge support from the JSPS Research Fellowship for Young Scientists, and R.K., T.A., K.O., and T.I. also thank the Graduate Program on Physics for the Universe (GP-PU) at Tohoku University for its support.
Part of the data analyzed in this study was acquired in the course of the Ph.D. work of Ryoko Kino at Tohoku University.

\bibliographystyle{apsrev4-2}
\bibliography{refs}

@article{Hiyama2001,
    author  = "Hiyama, E. and others",
    title   = "{$\Lambda$-$\Sigma$ conversion in $^{4}_Λ$He$ and ^{4}_Λ$H based on a four-body calculation}",
    journal = {Physical Review C},
    volume  = {65},
    pages   = {011301},
    year    = {2001},
    doi     = {10.1103/PhysRevC.65.011301},
    url     = {https://doi.org/10.1103/PhysRevC.65.011301}
}

@article{Fujiwara2007,
    author  = "Y. Fujiwara and Y. Suzuki and C. Nakamoto",
    title   = "{Baryon–baryon interactions in the $SU_6$ quark model and their applications to light nuclear systems}",
    journal = {Progress in Particle and Nuclear Physics},
    volume  = {58},
    pages   = {439--520},
    year    = {2007},
    doi     = {10.1016/j.ppnp.2006.08.001},
    url     = {https://doi.org/10.1016/j.ppnp.2006.08.001}
}

@article{Kamada1998,
    author  = "Kamada, H. and others",
    title   = "{Faddeev calculations of the hypertriton with modern nucleon--nucleon and hyperon--nucleon forces}",
    journal = {Physical Review C},
    volume  = {57},
    pages   = {1595--1606},
    year    = {1998},
    doi     = {10.1103/PhysRevC.57.1595},
    url     = {https://doi.org/10.1103/PhysRevC.57.1595}
}

@article{Lonardoni2017,
    author  = "Lonardoni, D. and Pederiva, F.",
    title   = "{Hyperon--nucleon interactions and hypernuclear structure from auxiliary field diffusion Monte Carlo}",
    journal = {Physical Review C},
    volume  = {95},
    pages   = {024003},
    year    = {2017},
    doi     = {10.1103/PhysRevC.95.024003},
    url     = {https://doi.org/10.1103/PhysRevC.95.024003}
}

@article{Garcilazo2015,
    author  = "Garcilazo, H. and Gal, A.",
    title   = "{Three-body calculations of $nn\Lambda$ and $np\Lambda$ systems}",
    journal = {Physical Review C},
    volume  = {92},
    pages   = {044606},
    year    = {2015},
    doi     = {10.1103/PhysRevC.92.044606},
    url     = {https://doi.org/10.1103/PhysRevC.92.044606}
}

@article{Haidenbauer2019,
    author  = "Haidenbauer, J. and others",
    title   = "{Hyperon--nucleon interaction at next-to-leading order in chiral effective field theory}",
    journal = {European Physical Journal A},
    volume  = {55},
    pages   = {23},
    year    = {2019},
    doi     = {10.1140/epja/i2019-12706-6},
    url     = {https://doi.org/10.1140/epja/i2019-12706-6}
}

@article{Garrido2024,
    author  = "E. Garrido and others",
    title   = "{Study of $p\Lambda$ and $pp\Lambda$ correlation functions with a three-body force tuned to hypertriton binding}",
    journal = {Physical Review C},
    volume  = {110},
    pages   = {054004},
    year    = {2024},
    doi     = {10.1103/PhysRevC.110.054004},
    url     = {https://doi.org/10.1103/PhysRevC.110.054004}
}

@article{Akaishi2026,
title = {Measurement of 3,4He(K-,p0)3,4H reaction cross section and evaluation of hypertriton binding energy},
journal = {Physics Letters B},
volume  = {873},
pages = {140163},
year = {2026},
issn = {0370-2693},
doi = {https://doi.org/10.1016/j.physletb.2026.140163},
url = {https://www.sciencedirect.com/science/article/pii/S0370269326000171},
author = {T. Akaishi and others}
}

@article{Audi2003,
    author = {G. Audi and A.H. Wapstra and C. Thibault},
    title = {The AME2003 atomic mass evaluation: II. Tables, graphs and references},
    journal = {Nuclear Physics A},
    volume = {729},
    pages = {337--676},
    year = {2003},
}

@article{Blomqvist1998,
    author = {K. I. Blomqvist and others},
    title = {The three-spectrometer facility at the Mainz microtron MAMI},
    journal = {Nuclear Instruments and Methods in Physics Research Section A: Accelerators, Spectrometers, Detectors and Associated Equipment},
    volume = {403},
    number = {2},
    pages = {263--301},
    year = {1998},
    issn = {0168-9002},
    doi = {10.1016/S0168-9002(97)01133-9},
    url = {https://doi.org/10.1016/S0168-9002(97)01133-9}
}

@article{Senger1993,
title = {The kaon spectrometer at SIS},
journal = {Nuclear Instruments and Methods in Physics Research Section A: Accelerators, Spectrometers, Detectors and Associated Equipment},
volume = {327},
number = {2},
pages = {393-411},
year = {1993},
issn = {0168-9002},
doi = {10.1016/0168-9002(93)90706-N},
url = {https://www.sciencedirect.com/science/article/pii/016890029390706N},
author = {P. Senger and others}
}

@article{Schulz2016NPA,
  title={\texorpdfstring{Ground-state binding energy of $^4_\Lambda$H from high-resolution decay-pion spectroscopy}{Ground-state binding energy of 4ΛH from high-resolution decay-pion spectroscopy}},
  author={F. Schulz and others},
  collaboration = {A1 Collaboration},
  journal={Nuclear Physics A},
  volume={954},
  pages={149--160},
  year={2016},
  publisher={Elsevier},
  doi = {10.1016/j.nuclphysa.2016.03.015},
  url = {https://doi.org/10.1016/j.nuclphysa.2016.03.015}
}

@article{Klag2018,
    author = {P. Klag and others},
    title = {Novel optical interferometry of synchrotron radiation for absolute electron beam energy measurements},
    journal = {Nuclear Instruments and Methods in Physics Research Section A},
    volume = {910},
    pages = {147--156},
    year = {2018},
    doi = {10.1016/j.nima.2018.09.072},
    url = {https://doi.org/10.1016/j.nima.2018.09.072}
}

@article{ajzenberg1988,
  title={Energy levels of light nuclei A= 5 - 10},
  author={Ajzenberg-Selove, Fay},
  journal={Nuclear Physics A},
  volume={490},
  number={1},
  pages={1--225},
  year={1988},
  publisher={Elsevier}
}

@article{cousins2008,
  title={Evaluation of three methods for calculating statistical significance when incorporating a systematic uncertainty into a test of the background-only hypothesis for a Poisson process},
  author={R.D. Cousins and others},
  journal={Nuclear Instruments and Methods in Physics Research Section A: Accelerators, Spectrometers, Detectors and Associated Equipment},
  volume={595},
  number={2},
  pages={480--501},
  year={2008},
  publisher={Elsevier}
}

@article{Wang2021AME2020,
  author       = {M. Wang and others},
  title        = {The AME 2020 atomic mass evaluation (II). Tables, graphs and references},
  journal      = {Chinese Physics C},
  year         = {2021},
  volume       = {45},
  number       = {3},
  pages        = {030003},
  doi          = {10.1088/1674-1137/abddaf},
  note         = {Atomic mass unit: NIST (2022 CODATA)}
}

@article{Kasagi2025,
      title={\texorpdfstring{Binding energy of $^{3}_{\Lambda}\mathrm{H}$ and $^{4}_{\Lambda}\mathrm{H}$ via image analyses of nuclear emulsions using deep-learning}{Binding energy of 3ΛH and 4ΛH via image analyses of nuclear emulsions using deep-learning}},
      author={A. Kasagi and others},     
    journal = {Progress of Theoretical and Experimental Physics},
    volume = {2025},
    number = {8},
    pages = {083D01},
    year = {2025},
    month = {07},
    issn = {2050-3911},
    doi = {10.1093/ptep/ptaf097},
    url = {https://doi.org/10.1093/ptep/ptaf097}
    }

@article{Juric1973,
  title={A new determination of the binding-energy values of the light hypernuclei (A<15)},
  author={M. Juri{\v{c}} and others},
  journal={Nuclear physics B},
  volume={52},
  number={1},
  pages={1--30},
  year={1973},
  publisher={Elsevier},
  url = {https://www.sciencedirect.com/science/article/pii/0550321373900849}
}

@article{Shao2022,
  title={\texorpdfstring{Measurement of $^4_\Lambda$H and $^4_\Lambda$He binding energy in Au+ Au collisions at $\sqrt{s_{NN}}= 3$~GeV}{Measurement of H Lambda 4 and He Lambda 4 binding energy in Au+ Au collisions at s NN= 3 GeV}},
  collaboration = {The STAR Collaboration},
  author={M.S. Abdallah and others},
journal = {Physics Letters B},
volume = {834},
pages = {137449},
year = {2022},
issn = {0370-2693},
publisher={Elsevier},
doi = {10.1016/j.physletb.2022.137449},
url = {https://doi.org/10.1016/j.physletb.2022.137449}
}

@article{STAR2020,
  title={Measurement of the mass difference and the binding energy of the hypertriton and antihypertriton},
  author = {J. Adam and others},
  collaboration = {The STAR Collaboration},
  journal={Nature Physics},
  volume={16},
  number={4},
  pages={409--412},
  year={2020},
  publisher={Nature Publishing Group UK London},
  doi = {10.1038/s41567-020-0799-7},
  url = {https://doi.org/10.1038/s41567-020-0799-7}
}

@misc{eckert2021chart,
  title={Chart of hypernuclides—Hypernuclear structure and decay data},
  author={P. Eckert and P. Achenbach and others},
  year={2021},
  url = {https://hypernuclei.kph.uni-mainz.de}
}

@article{le2020,
  title={\texorpdfstring{Implications of an increased $\Lambda$-separation energy of the hypertriton}{Implications of an increased Lambda-separation energy of the hypertriton}},
  author={H. Le and others},
  journal={Physics Letters B},
  volume={801},
  pages={135189},
  year={2020},
  publisher={Elsevier}
}

@article{Hiyama2014,
  title={\texorpdfstring{Three-body model study of $nn\Lambda$ with realistic $YN$ and $NN$ interactions}{Three-body model study of nnΛ with realistic YN and NN interactions}},
  author={E. Hiyama and others},
  journal={Physical Review C},
  volume={89},
  number={6},
  pages={061302},
  year={2014},
  publisher={American Physical Society},
  doi = {10.1103/PhysRevC.89.061302},
  url = {https://doi.org/10.1103/PhysRevC.89.061302}
}

@article{Acharya2023,
  title = {\texorpdfstring{Measurement of the Lifetime and $\Lambda$ Separation Energy of $^3_\Lambda$H}{Measurement of the Lifetime and Lambda Separation Energy of 3ΛH}},
  author = {S. Acharya and others},
  collaboration = {The ALICE Collaboration},
  journal = {Physical Review Letters},
  volume = {131},
  issue = {10},
  pages = {102302},
  numpages = {13},
  year = {2023},
  month = {9},
  publisher = {American Physical Society},
  doi = {10.1103/PhysRevLett.131.102302},
  url = {https://link.aps.org/doi/10.1103/PhysRevLett.131.102302}
}

@article{Bohm1968,
title = "A determination of the binding-energy values of light hypernuclei",
journal = "Nuclear Physics B",
volume = 4,
number = 6,
pages = "511--526",
year = 1968,
doi = "10.1016/0550-3213(68)90109-0",
url = "www.sciencedirect.com/science/article/pii/0550321368901090",
author = "G. Bohm and others"}

@article{Gajewski1967,
title = "A compilation of binding energy values of light hypernuclei",
journal = "Nuclear Physics B",
volume = 1,
number = 3,
pages = "105--113",
year = 1967,
doi = "10.1016/0550-3213(67)90095-8",
url = "www.sciencedirect.com/science/article/pii/0550321367900958",
author = "W. Gajewski and others"}

@article{DAVIS1992369,
title = {Hypernuclei-the early days},
journal = {Nuclear Physics A},
volume = {547},
number = {1},
pages = {369-378},
year = {1992},
issn = {0375-9474},
doi = {https://doi.org/10.1016/0375-9474(92)90746-7},
url = {https://www.sciencedirect.com/science/article/pii/0375947492907467},
author = {D.H. Davis}
}

@article{ALICE:2024djx,
author = "S. Acharya and others",
collaboration = "The ALICE Collaboration",
title = "{First Measurement of $A$ = 4 Hypernuclei and Antihypernuclei at the LHC}",
doi = "10.1103/PhysRevLett.134.162301",
journal = "Physical Review Letters",
volume = 134,
number = 16,
pages = 162301,
year = 2025}

@article{PDG2024,
  author = {S. Navas and others},
  collaboration = {Particle Data Group},
  title = {Review of Particle Physics},
  journal = {Physical Review D},
  volume = {110},
  pages = {030001},
  year = {2024},
  doi = {10.1103/PhysRevD.110.030001}
}

@article{bohm1970investigation,
  title={An investigation of the range-energy relation in emulsion},
  author={G. Bohm and others},
  journal={Il Nuovo Cimento A (1965-1970)},
  volume={70},
  number={3},
  pages={384--390},
  year={1970},
  publisher={Springer}
}

@article{chen2023,
  title={Measurements of the lightest hypernucleus (H$\Lambda$3): progress and perspective},
  author={J. Chen and others},
  journal={Science Bulletin},
  volume={68},
  number={24},
  pages={3252--3260},
  year={2023},
  publisher={Elsevier}
}

\end{document}